\documentclass{elsarticle}

\usepackage{mathrsfs}
\usepackage{lineno,hyperref}
\usepackage{tikz}
\usepackage[export]{adjustbox}
\usepackage{amsmath}
\usepackage{geometry,tikz}
\usetikzlibrary{arrows,matrix,positioning,fit}
\usepackage{amssymb}
\usepackage{physics}
\usepackage{float}
\usepackage{subcaption}
\usepackage{array}
\usepackage{multirow}
\usepackage{relsize}
\usepackage{ulem}
\usepackage{enumitem}
\modulolinenumbers[5]

\journal{Applied Ocean Research}

\newcommand{\fig}[1]{Fig.~\ref{#1}}
\newcommand{\eq}[1]{Eq.~(\ref{#1})}
\newcommand{\eqp}[1]{Eq.~\ref{#1}}

\newcommand{\ga}{\gtrsim}
\newcommand{\la}{\lesssim}
\newcommand{\rom}[1]{\textcolor{black}{#1}}

\DeclareCaptionFormat{cont}{#1 #2#3\par}

\newcommand{\vz}{w} 
\newcommand{\sx}{\eta_{,x}}

\newcommand{\sw}{{\sigma_{\vz}}}
\newcommand{\ssx}{{\sigma_{\sx}}}
\newcommand{\hs}{H_{\rm s}}
\newcommand{\tz}{T_{\rm z}}

\newcommand{\kst}{\kappa}
\newcommand{\Dcum}{D_{\rm cum}}
\newcommand{\nipst}{{N_{\rm ip}}^{\rm (st)}}
\newcommand{\nlt}{N_{\rm lt}}
\newcommand{\nrea}{N_{\rm rea}}
\newcommand{\nreaPlot}{20}
\newcommand{\tvib}{t_{\rm vib}}
\newcommand{\tip}{t_{\rm imp}}
\newcommand{\dlt}{d_{\rm lt}}
\newcommand{\dst}{d_{\rm st}}
\newcommand{\psmall}{P_{\varepsilon}}
\newcommand{\pfail}{P_{\rm fail}}
\newcommand{\pfailu}{{\pfail}^{({\rm u})}}
\newcommand{\sn}{$SN$}

\newcommand{\Su}{S_{\rm u}}
\newcommand{\Se}{S_{\rm e}}
\newcommand{\Sz}{S_{\rm 0}}
\newcommand{\sst}{S_{*}}
\newcommand{\sstu}{{S_{*}}^{({\rm u})}}
\newcommand{\scyU}{s^{(1)}}
\newcommand{\scyK}{s^{(k)}}
\newcommand{\epm}{\epsilon_{\rm m}}
\newcommand{\za}{z_{\rm a}}
 \newcommand{\distDir}{\mathcal{A}}
\newcommand{\pcs}{P_{\rm cs}}
\newcommand{\acs}{\alpha_{\rm cs}}
\newcommand{\bcs}{\beta_{\rm cs}}
\newcommand{\gcs}{\gamma_{\rm cs}}
\newcommand{\sigcs}{\sigma_{\rm cs}}
\newcommand{\mucs}{\mu_{\rm cs}}
\newcommand{\lc}{a}

\begin{document}

\begin{frontmatter}

\title{On the risk of fatigue failure of structural elements
exposed to bottom wave slamming -- Impulse response regime}

% %%%%%%%%%%%%%%%%%%%%%%%%%%%%%%%

\author[enstaAddress]{Romain Hasco\"{e}t\corref{cor1}}
\cortext[cor1]{Corresponding author}
\ead{romain.hascoet@ensta-bretagne.fr}

\author[enstaAddress]{Nicolas Jacques}

\address[enstaAddress]{ENSTA Bretagne, CNRS UMR 6027, IRDL, 2 rue Fran\c{c}ois Verny, 29806 Brest Cedex 9, France}

\begin{abstract}

This study aims to investigate whether 
%the 
fatigue damage induced by bottom wave slamming
can be a failure mode, important to consider when sizing a marine structural element.
The body exposed to wave impacts is assumed to have a shape and structural arrangement such that
the duration of wave-impact loads is short relative
to the structure's vibratory response time.
In this dynamical regime, fatigue is found to be a potentially
important failure mechanism: 
accounting for the risk of failure due to fatigue damage 
may result in design constraints that 
are significantly more conservative than those based on
the risk of ultimate strength exceedance.
The role of fatigue damage depends on the elevation of the body.
It is predominant for low elevations, for which slamming events are frequent.
Since this study aims to provide general \rom{insight},
the specific details of the body, such as its shape and structural arrangement,
are not specified.
Instead, a general framework is used for the analysis.
The way forward to address a specific case study, possibly including the effects of forward and seakeeping motions, is briefly explained.

\end{abstract}


\end{frontmatter}

\section{Introduction}

Fatigue damage can be a critical failure mechanism to consider 
in the design of ships and offshore 
structures.
One of the 
primary environmental factors causing time-varying mechanical stress,
which leads to fatigue damage, is 
exposure to water waves.
Water-wave loads can induce time-varying stress through 
several
mechanisms: 
(i) non-resonant response (possibly quasistatic) to the continuous 
wave loading
(see, e.g., \cite{jha_2000, li_2013, li_2014}), 
(ii) resonant response to the continuous 
wave loading
(\textit{springing} mechanism, see, e.g.,
\cite{naess_1992, winterstein_1994, storhaug_2007b}),
(iii) transient events related to high-order wave nonlinearities (see, e.g., \cite{faltinsen_1995, tromans_2006, bachynski_2014}),
(iv) transient events induced by wave slamming (see, e.g., \cite{wienke_2005, barhoumi_2014, hulin}).
Much research has been devoted to the experimental investigation
and the modeling of the fatigue of ships, 
driven
by the global response of their hull girders to wave loads
(see \cite{temarel_2016, dong_2022} for recent reviews).
Ship hull girders
are subject to wave-loading mechanisms 
\rom{(i-ii-iv).}
%(i), (ii), (iv).
Wave slamming is usually regarded as a secondary 
source of fatigue damage (see, e.g.,  \cite{mao_2010b, andersen_2013}), 
acting through the global stress-response of the hull girder (\textit{whipping} mechanism).
In contrast, 
to our knowledge, configurations where wave-slamming 
is the primary
source of fatigue damage have not been considered in the literature.
The latter situation may occur for structural elements, 
either on ships or offshore installations, 
that respond locally to wave slamming.

The objective of this
paper is to investigate whether fatigue damage could 
be a failure mode relevant to the design of a marine 
structural element, 
whose main source of stress variations
is bottom wave slamming.
The exposed body is assumed to be above the still water level, at a 
certain elevation.
If the body is close to the mean sea level, it
will frequently experience 
water entry and emersion cycles.
If its elevation is increased, the body
will remain mostly out of the water, 
with wave impacts occurring less frequently.
In both cases,
the exposed body will experience bottom-slamming loads due to the successive water entries.
In this context, slamming events are defined as 
the moments
when the body crosses the sea free-surface,
into the water domain. 
It is important to note
that even moderate water waves \rom{may} generate substantial
slamming loads (especially for blunt bodies), 
since 
the structural element is assumed to enter the water domain from a fully emerged position.
The novelty
of this study lies in the consideration 
that the local response of the structural element to bottom-wave slamming
may constitute the dominant contribution to fatigue damage. 

\textit{A priori}, the modeling of the risk of failure of a structure subjected to wave impacts
 requires considering 
 two potential failure modes: 
\begin{itemize}
\item Failure due to 
stress locally exceeding a specific threshold. The considered threshold is usually the ultimate tensile strength 
(shortened hereinafter to ``ultimate strength'')
or the yield strength of the material.
The choice between the ultimate and yield strengths depends on whether 
the occurrence of localized 
plastic deformation is acceptable.
When the structural element is designed for a low probability of failure,
this failure mode relates to extreme events (i.e., extreme wave impacts)
that
have a low probability of occurrence over the considered exposure time.
\item Failure due to fatigue damage. 
Fatigue damage will 
accumulate through a (very) 
large number of impacts, experienced
over the exposure time.
\end{itemize}
A key point
in modeling the first failure mode is 
apprehending rare and extreme events. 
This presents challenges in two regards.
Firstly, it requires modeling
the probability of occurrence of extreme environmental conditions: 
here, extreme waves, which are themselves generated by extreme sea states.
Second, the response of the system to these extreme conditions needs to be estimated,
with the substantial challenge posed by the fact
that the kinematics of extreme waves remains uncertain.

The second failure mode (fatigue damage) may be less challenging in terms of excitation and response modeling,
as most 
of the damage may be 
caused by a class of  events that 
are not particularly rare and extreme.
For such a class of events, in situ or lab measurements may be ``readily'' accessible,
while modeling is less complicated and less subject to uncertainties.
However, to properly model this second failure mode,
a key challenge
lies in the fact that the long-term fatigue damage 
results from
the accumulation of a large number of random events (here wave impacts),
whose underlying probability distribution (here related to the properties of the encountered sea states) is also random.
In this respect, this makes `risk-based' analyses (i.e., the sizing of the marine structure for a given probability of failure) 
more involved than for the first failure mode.

This
paper investigates the relative importance of these two failure modes,
when designing a structural element with respect to the risk induced by bottom wave slamming. 
The configuration is assumed to be such that the structural element experiences a 
large number of wave impacts over its lifetime.
The case where the risk of wave impact occurrence is small over the exposure time
 is not relevant to fatigue damage.
 The latter situation would typically arise
 when the elevation of the exposed body 
(vertical position of the body relative to the still water level) 
 is significantly larger 
 than the typical significant wave height 
 in the geographical zone of interest
 (see, e.g., \cite{sweetman_2003, lim_2019}).
 As the relative importance of the two failure modes depends on the total number of experienced wave impacts,
 this study explores how
 the results vary with
 the assumed exposure duration and body elevation. 
 Additionally, it aims to identify the dominant classes of sea states and wave impacts,
that contribute most to the risk.
The study adopts a general framework, 
allowing the findings and conclusions to remain qualitatively applicable across various contexts.
The structural element exposed to wave slamming is assumed to have a 
bottom shape and a structural arrangement such that the characteristic 
duration of slamming loads
is short 
compared to the structure's response timescale. 
The local vibratory stress response is assumed to be dominated by a single structural mode.
This dominant vibratory mode is assumed to be responsible 
for most of the fatigue damage.

The framework considered in the present study is further 
elaborated
in Section \ref{sec_framework}.
It details the assumptions 
regarding the structural stress response following wave impact (\S\ref{subsec_dyn_regimes}) and
the resulting fatigue damage (\S\ref{subsection_fatigue_model}). 
Additionally, it outlines the assumptions related to the modeling of sea state occurrence (\S\ref{subsec_model_st_occurrence}), 
and impact occurrence within
a given sea state (\S\ref{subsec_model_ip_occurrence}).
Section \ref{sec_case_studies} introduces the case studies to be considered for investigation and discussion.
Section \ref{sec_prob_fail} investigates the evolution of the failure probability as a function of the structural sizing,
exposure time, and failure mode.
The dominant classes of sea states and wave impacts 
responsible for the risk of fatigue failure
are 
identified in Section \ref{sec_rea_to_fail}.
The effect of a change in the elevation of the exposed body is investigated in Section \ref{subsec_effect_aLevel}.
The paper ends with a discussion in Section \ref{sec_discussion}, 
followed by a conclusion in Section \ref{conclusion}.
 
\vspace{-0.2cm}
\section{Framework and Assumptions}
\label{sec_framework}
\vspace{-0.05cm}

A flow diagram summarizing the 
approach is presented in
Fig. \ref{fig_flow_diagram}.
Each component of the methodology
is detailed below.

% %%%%%%%%%%%%%%%%%%%%%%%%
% %%%%%%%%%%%%%%%%%%%%%%%%
% %%%%%        FIGURE    BEGIN         %%%%%
% %%%%%%%%%%%%%%%%%%%%%%%%
% %%%%%%%%%%%%%%%%%%%%%%%%

\def\scaleF{0.8}
\def\fI{}

\begin{figure}[h!]
\begin{center}
\begin{tabular}{c} 
         \includegraphics[width=\scaleF\textwidth]{\fI 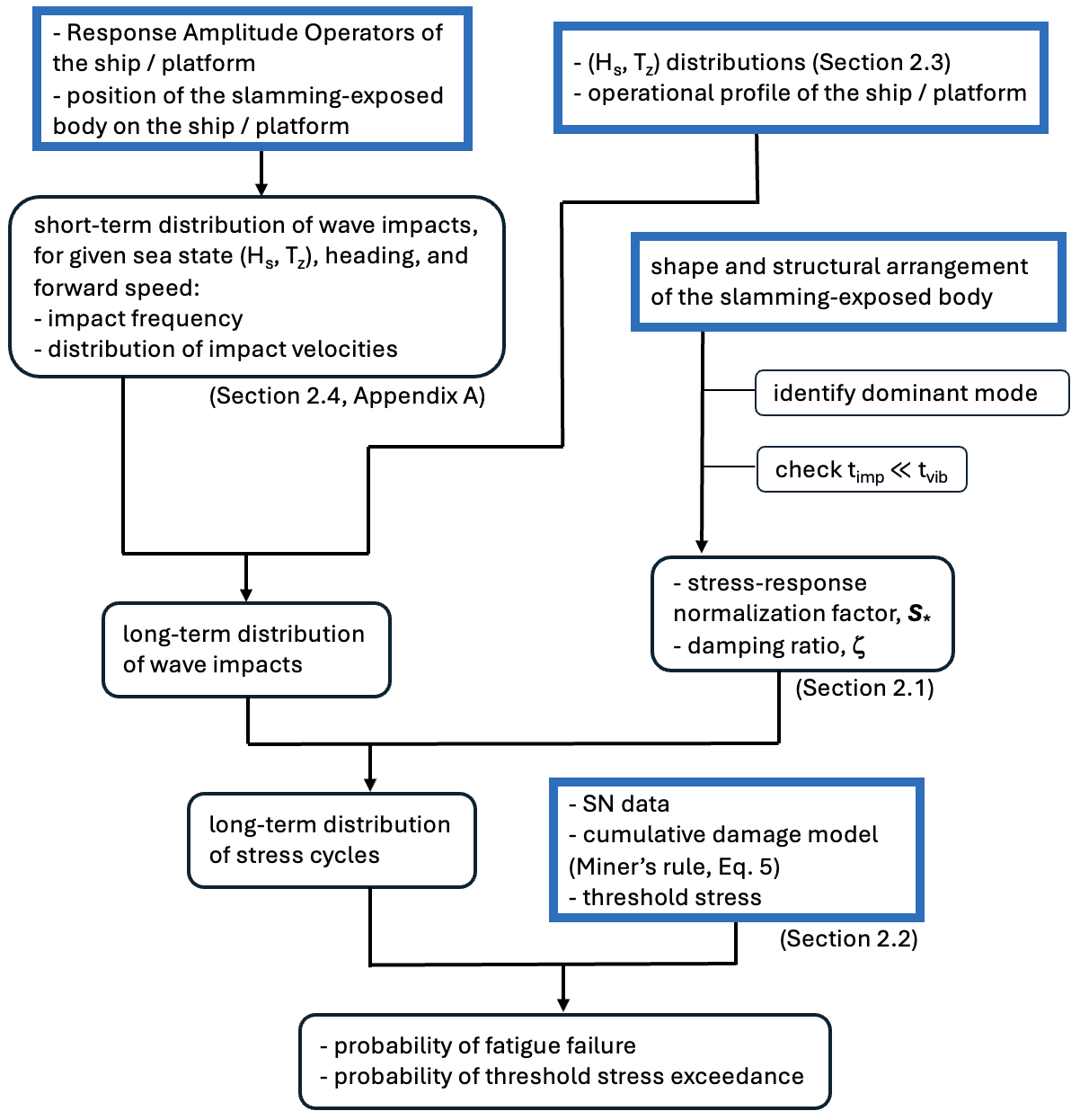}
\end{tabular}        
\end{center}
\caption{
Flow diagram of the approach. 
The input data of the problem are 
indicated by blue boxes with square corners.
}
\label{fig_flow_diagram}
\end{figure}

% %%%%%%%%%%%%%%%%%%%%%%%%
% %%%%%%%%%%%%%%%%%%%%%%%%
% %%%%%        FIGURE    END         %%%%%
% %%%%%%%%%%%%%%%%%%%%%%%%
% %%%%%%%%%%%%%%%%%%%%%%%%

\vspace{-0.1cm}
\subsection{Structural response to wave impacts: impulse regime}
\label{subsec_dyn_regimes}

During a hydrodynamic impact on an elastic body, 
transient loads will induce a vibratory response of the body structure,
provided that the loads evolve sufficiently quickly
relative to
the structure's response timescale 
(period of the dominant mode)
-- see \cite{faltinsen_2006, faltinsen_2000, korobkin_1999, scolan_2004, sun_2006, maki_2011, piro_2013} 
for a sample of studies on slamming and the resulting hydroelastic response.
In the context of 
bottom-wave slamming,
the vertical component of the fluid at impact may be considered as the main variable relevant to 
slamming loads (see, e.g., \cite{ochi_1973, rassinot_1995, wang_2002, hermundstad_2007, wang_2016}).
The impact timescale, $\tip$, 
can be expressed as follows:
\begin{equation}
\label{eq_tip}
\tip = \frac{h_{\rm imp}}{V} \, ,
\end{equation}
where $h_{\rm imp}$ is the characteristic water-entry depth over which the transient loads develop, and $V$ 
is the vertical component of the characteristic water-entry velocity.
It is important to note
that $h_{\rm imp}$ may be significantly smaller than the size of the exposed body, particularly for blunt-shaped bodies, 
such as a tubular element of circular section (see, e.g., \cite{cointe_1987}), or a water foil (see, e.g., \cite{hascoet_2019}).
In this study, 
the response timescale of the body's structure, $\tvib$, is assumed to be 
much longer than $\tip$:\footnote{
\label{footnote_harmonic}
In practice, results obtained for a simple harmonic oscillator suggest that $\tvib/\tip \ga 3$ may be sufficient.} 
\begin{equation}
\label{eq_condi_ton_tvib}
\tvib \gg \tip \, .
\end{equation}
Then, as illustrated in \fig{fig_sketch}, the slamming load may be 
treated
as an impulse from the perspective of structural response.
Hereafter, this dynamical regime is 
referred to as the \textit{impulse regime}.
Faltinsen (1999) \cite{faltinsen_1999} explored this impulse regime
and the transition to the non-impulse regime,
through numerical simulations of wedge-shaped structures 
entering water at constant velocity.

Considering a structural detail within
the body,
the impact-induced vibratory response will translate into a 
series of stress cycles
that may induce fatigue damage.
For simplicity,
a single structural mode is assumed 
to dominate the vibratory response, as well as the resulting fatigue damage: 
hereafter, $\tvib$ is used to refer to
the eigen period of the dominant mode.
Note that the dominant mode is not necessarily the mode having the lowest eigen frequency,
but the mode that leads to the largest stress amplitude (for the considered loading).
Hydroelastic coupling must
be considered when evaluating $\tvib$, due to added-inertia effects.

% %%%%%%%%%%%%%%%%%%%%%%%%
% %%%%%%%%%%%%%%%%%%%%%%%%
% %%%%%        FIGURE    BEGIN         %%%%%
% %%%%%%%%%%%%%%%%%%%%%%%%
% %%%%%%%%%%%%%%%%%%%%%%%%

\def\scaleF{0.28}
\def\fI{}

\begin{figure}[t!]
\begin{center}
\begin{tabular}{c} 
	$\tvib \gg \tip$ \\
         \includegraphics[width=\scaleF\textwidth]{\fI 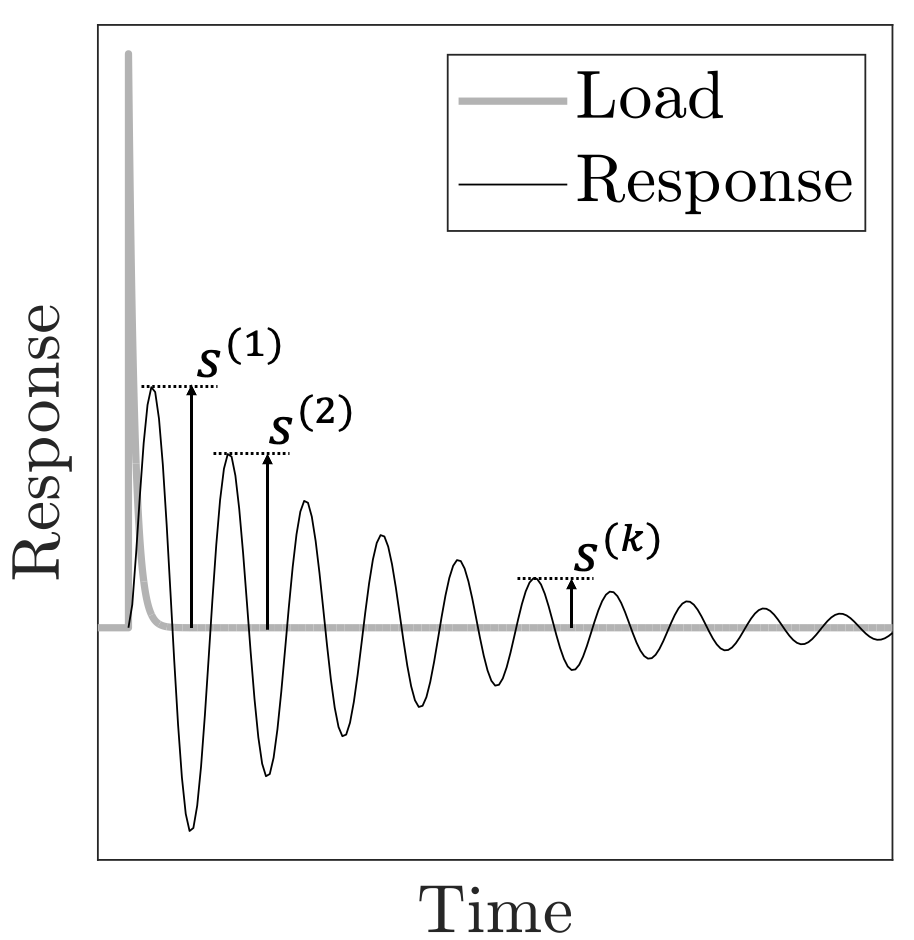}
\end{tabular}        
\end{center}
\caption{
Illustration of the impulse stress-response regime considered in the present study.
The stress response is assumed to be dominated by a single structural mode.
In this dynamical regime, 
the stress amplitude of the first cycle, $\scyU$, is approximately proportional to the load impulse 
(i.e., the time-integral of the load). 
}
\label{fig_sketch}
\end{figure}

% %%%%%%%%%%%%%%%%%%%%%%%%
% %%%%%%%%%%%%%%%%%%%%%%%%
% %%%%%        FIGURE    END         %%%%%
% %%%%%%%%%%%%%%%%%%%%%%%%
% %%%%%%%%%%%%%%%%%%%%%%%%

Following the impulse excitation, if 
there are no further rapid changes
in the hydrodynamic load,
the mechanical vibrations will decay. 
The damping of the vibratory response is assumed to be of viscous type (see, e.g., \cite{schmitz_2012}),
which is typically
valid for metal structures.
For simplicity, 
the damping ratio of the dominant mode, $\zeta$, is assumed to be constant and 
significantly smaller than unity, $\zeta\ll1$.
The assumption $\zeta\ll1$ is usually valid for metal structures in air \cite{schmitz_2012}.
In the context of hull structures, measured values of damping ratios are typically around $1\%$
(see, e.g., \cite{storhaug_2007c, storhaug_2014, orlowitz_2014, vanzijl_2021}).
In the specific context of structural elements exposed to wave slamming, 
the main uncertainty may lie in the potential effects of hydroelastic coupling on the damping ratio.
This subject has received limited
attention in the literature 
(see \cite{abrahamsen_2023} for a recent study addressing this matter).
Several experimental studies (see, e.g., \cite{campbell_1980, sujathauvin_2017, wang_2020, moalemi_2023})
show that water impacts do generate a vibratory structural response.
Although the damping ratio is not evaluated in these studies, it is clearly smaller than unity.
Moreover, in lab experiments, the measurement of impact forces 
often requires a special treatment 
to compensate for the perturbations induced by the vibrations of the mock-up \cite{antonini_2021, tassin_2024}.

Based on the aforementioned assumptions 
regarding structural damping,
the stress response of a structural detail, 
induced by a wave impact, can be modeled as a series
of cycles with exponentially decaying amplitudes, 
which can be approximated as follows: 
\begin{equation}
\label{eq_sn_s1}
\frac{\scyK}{\scyU} \simeq \exp\{ - 2 \pi \zeta (k-1) \} \, ,
\end{equation}
where $\scyU$ and $\scyK$ represent
the amplitudes of the $1^{\rm st}$ and $k^{\rm th}$ cycles, respectively
(see \fig{fig_sketch} for an illustration).
The structural vibratory response is assumed to have time to mostly 
dampen between 
%two 
successive wave impacts.

According to \eq{eq_sn_s1}, 
the sequence of stress-cycle amplitudes, triggered by a wave impact, 
is entirely determined by the amplitude of the first cycle, $\scyU$.
In the present study, 
the magnitude of the stress response is assumed to depend solely on the vertical component of the fluid velocity.\footnote{
Depending on the shape of the slamming-exposed body,
and the desired degree of accuracy, 
it may be important to include additional kinematic variables 
-- 
such as the free-surface slope, the horizontal fluid velocity, or the fluid acceleration -- 
as inputs to the considered 
slamming model (see, e.g., \cite{scolan_2015, helmers_2012, hascoet_2019, hascoet_2020, hascoet_2021}).
}
The slamming impulse, which is the time integral 
of the slamming force,
responds linearly to the impact velocity.
This linear relationship arises from two factors:
(i) slamming loads are proportional to the square of the water-entry velocity
(if the latter is assumed to be constant), and
(ii) the characteristic water-entry duration, 
as defined by \eq{eq_tip},
is inversely proportional to the impact velocity.
Then, if the stress responds linearly to the load impulse
(which is expected if condition (\ref{eq_condi_ton_tvib}) is fulfilled 
and if the structure has a linear elastic behavior),
the amplitude of the first cycle may be modeled as follows:
\begin{equation}
\label{eq_s1_ss} 
\displaystyle \scyU = \sst \left( \frac{V}{1 \ {\rm m/ s}} \right) \, ,
\end{equation}
where $\sst$ is 
a stress normalization factor 
that depends on the sizing of the structural element 
under consideration.
This
study focuses on the relative importance of the two aforementioned failure modes 
as a function of $\sst$,
without specifying the underlying structural element.
The effective estimation of $\sst$ for a specific structural element 
is illustrated in \ref{fit_sstart_faltinsen}.

\subsection{Fatigue damage and failure criteria}
\label{subsection_fatigue_model}

This subsection 
details the approach used to model fatigue damage.
In \S\ref{subsubsec_snpattern} the assumed \sn{} curve pattern is introduced.
Section \ref{subsubsec_snRnd} then explains how the randomness of the \sn{} curve is modeled.
Finally, \S\ref{subsubsec_fail_crit} briefly discusses the failure criteria 
considered in this study.

\subsubsection{\sn{} curve pattern}
\label{subsubsec_snpattern}

Various empirical laws have been proposed to model the fatigue stress-life 
relationships of metals 
(see \cite{bannantine_1990,stephens_2000} for reviews on the subject).
These relationships are commonly known
as ``\sn{} curves'', where $N$ 
represents the number of cycles to failure
at a given stress-cycle amplitude (or range), $S$.
The metal considered in the present marine context is steel, as it is widely used in the naval and offshore industries.
For fatigue damage occurring
in the high cycle regime ($10^7>N>10^4$) 
or in the very high cycle regime ($N>10^7$), 
classification societies provide
guidelines regarding the \sn{} curve patterns %%%, which can 
to be used for the fatigue assessment of ship structures (see, e.g., \cite{bv_2020, dnv_2021a, abs_2017}).
These \sn{} curve patterns are composed
of two power-law branches: 
\begin{itemize}
\item A high cycle power-law branch where $S \propto N^{-1/m}$, 
which is valid for $10^7>N>10^4$.  
Classification societies recommend 
an index $m$ ranging from $3$ to $4$, 
depending on the geometry and surface finishing of the structural detail.
The value $m=3$ is 
used in the illustrative examples investigated below. 
\item A very high cycle power-law branch where $S \propto N^{-1/(m+\Delta m)}$, starting from $N = 10^7$, 
with $\Delta m = 2$ being the value recommended by classification societies.
The two branches connect at $N=10^7$.
In order to limit the number of cycles that need to be considered, 
the very-high cycle branch is truncated above $N=10^{12}$;
i.e., an endurance limit is set to $S(N=10^{12})$.
This truncation is adopted for 
computational convenience
and does not affect the qualitative findings of this study.
\end{itemize}
Through the illustrative examples provided
in Section \ref{sec_results},
it will be shown that this \sn{} curve pattern 
-- limited to $N>10^4$ --
is sufficient to estimate the risk of fatigue failure, 
provided the exposed body is 
close enough to the mean water level.
This latter condition ensures a large number of wave impacts over the body's life.
However, as the body's elevation 
above the mean water level increases,
the frequency of wave impacts decreases, 
and at a certain
point, low cycle fatigue becomes relevant
in the reliability analysis
(this point is specifically addressed in \S\ref{subsec_effect_aLevel}).
Furthermore,
in the following sections, 
the risk induced by fatigue is 
compared with the risk of failure due to 
the material's ultimate strength being exceeded.
For these two reasons, the 
risk analysis must account for two additional factors:
(i) the damage contribution of stress cycles within the low cycle regime;
(ii) the possibility that a single extreme event leads to failure through the 
exceedance of the ultimate strength.
In order to cover these two aspects, the \sn{} curve pattern proposed 
by classification societies has been extended into 
the low cycle region ($N<10^4$) by introducing a third power-law branch. 
To decide on how to parametrize this third branch, 
the ultimate strength of the material, $\Su$, has been considered to be a good proxy 
for the stress cycle amplitude at $N=1$, i.e., $S(N=1) \equiv \Sz \simeq \Su$.\footnote{
An alternative option could have been to map $\Su$ 
to $S(N=1/2)$ or $S(N=1/4)$; 
$N=1/2$ and $N=1/4$ respectively corresponding to a half-cycle and a single monotonic tensile ramp:
the precise choice is unimportant in the present study.
}
On the other hand, the yield strength of the material, $\Se$, may be considered as a good proxy for
the transition between the low cycle and high cycle regimes, i.e., $S(N=10^4) \simeq \Se$. 
Given that most steel types used in the marine industry typically have $\Su/\Se \simeq 2$ (see, e.g., \cite{bv_2021, dnv_2021c, abs_2022}),
these considerations have led to the choice $\Sz = 2 \times S(N=10^4)$.
These assumptions produce
the \sn{} curve pattern shown in 
the left panel of \fig{fig_SNcurve_pattern}.

% %%%%%%%%%%%%%%%%%%%%%%%%
% %%%%%%%%%%%%%%%%%%%%%%%%
% %%%%%        FIGURE    BEGIN         %%%%%
% %%%%%%%%%%%%%%%%%%%%%%%%
% %%%%%%%%%%%%%%%%%%%%%%%%

\def\scaleF{0.45}

\begin{figure}[h!]
\begin{center}
\begin{tabular}{cc} 
	\includegraphics[width=\scaleF\textwidth]{\fI 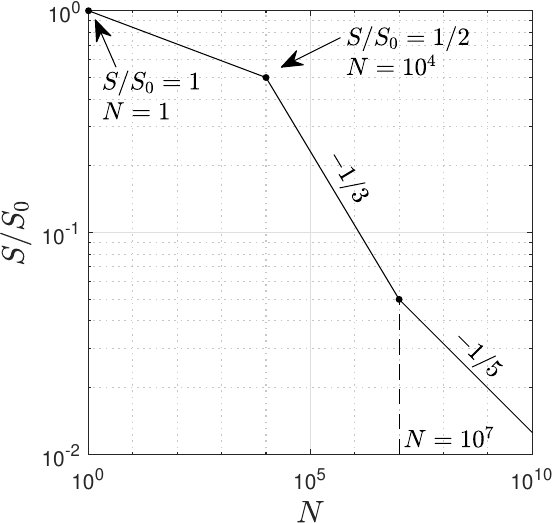}  & 
	\includegraphics[width=\scaleF\textwidth]{\fI 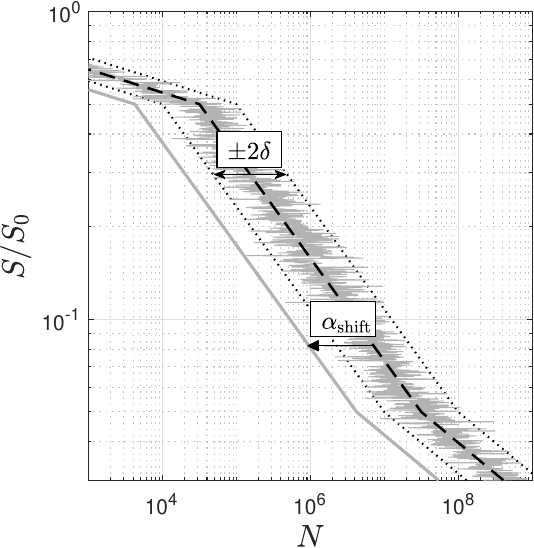} 
\end{tabular}        
\end{center}
\caption{
\textit{Left:} 
\sn{} curve pattern used in this
study. The different constraints which define this \sn{} curve are annotated:
(i) the stress amplitude at $N=1$, $S_0 \equiv S(N=1)$, is used as a normalization factor of the proposed $\sn{}$ curve pattern. 
$S_0$ is expected to be close to the ultimate strength of the material;
(ii) the stress level at the end of the low cycle branch is set to $S=\Sz/2$ at $N=10^4$.
The transition stress level, $\Sz/2$, can be considered as a proxy for the yield strength;
(iii) the power law index of the high cycle branch is equal to $-1/3$;
(iv) the very high cycle branch starts at $N= 10^7$ and has a power law index equal to $-1/5$.
\textit{Right:} \sn{} curve randomness. 
In the illustration, the median \sn{} curve is shown as a dashed line. 
The noisy \rom{gray} curve illustrates the case 
where the randomness is modeled through option 2 (see \S\ref{subsubsec_snRnd})
The solid \rom{gray} curve illustrates the case where the randomness is modeled through 
a random shift factor, $\alpha_{\rm shift}$ (option 1). 
The dotted lines show the $\pm 2\delta$ band. 
The ``$- 2\delta$'' curve (i.e., the dotted line on the left of the median \sn{} curve) matches the curve pattern shown on the left panel.
}
\label{fig_SNcurve_pattern}
\end{figure}

% %%%%%%%%%%%%%%%%%%%%%%%%
% %%%%%%%%%%%%%%%%%%%%%%%%
% %%%%%        FIGURE    END         %%%%%
% %%%%%%%%%%%%%%%%%%%%%%%%
% %%%%%%%%%%%%%%%%%%%%%%%%

The mean stresses of cycles are expected to be close to zero (full stress reversal) in the impulse regime.
Therefore, the effect of mean stress on fatigue life does not need to be considered in this study. 

\subsubsection{\sn{} curve randomness}
\label{subsubsec_snRnd}

Fatigue tests show significant
dispersion in the results obtained from experiments 
repeated in the same conditions. 
For a given fixed stress amplitude, 
the observed number of cycles to failure can vary 
considerably between trials.
As stated by Stephens et al. (2000) \cite{stephens_2000} this variability may 
arise from ``inherent material variability, 
variations in heat treatment and manufacturing, variations in specimen or component geometry,
and variations from differences in the testing conditions''.
In this study,  
the structural detail under study is linked to an \sn{} curve, $n(S)$,
which is modeled as a random curve.
For a given structural detail specimen, the realized instance of 
the \sn{} curve is assumed to hold during the whole life of the specimen.
Then, the fatigue damage, $\Dcum$, accumulated by the specimen is computed 
using Miner's rule \cite{miner_1945}.
Assuming that the stress loading history consists of a collection of 
$M$ cycles of amplitudes $s_1$, $s_2$, ..., $s_M$,
Miner's rule reads:
\begin{equation}
\label{eq_dcum}
\Dcum = \sum_{k=1}^{M} \frac{1}{n(s_k)} \, .
\end{equation}
For a given stress amplitude, the statistical properties of $n(S)$, 
can be estimated by conducting
fatigue tests at different constant stress amplitudes,
using a different specimen for each test.
However, for a realized \sn{} curve (i.e., for a given specimen) 
the covariance structure between $n(S_1)$ and $n(S_2)$, 
the numbers of cycles to failure at two different stress levels, $S_1$ and $S_2$, is more difficult to probe, 
as a fatigue test cannot be repeated twice with the same specimen.
In the literature, this matter has been investigated by modeling $n(S)$ 
as the realization of a stochastic process (see, e.g., \cite{liu_2007, paolino_2014}).
For the present study, two extreme assumptions 
have been tested:

\begin{enumerate}
\item 
The \sn{} curve is randomized through a global shift factor applied 
to the median \sn{} curve, $N(S)$.
In other words, for a given specimen, the knowledge of $n(S)$ at a given stress level fully determines the realized \sn{} curve. 
This approach corresponds to the one proposed by classification societies \cite{bv_2020, dnv_2021a}.
In this study, the logarithm of the random shift factor 
is assumed to follow a centered normal distribution of standard deviation $\delta_{\rm BV} = 0.25\times \ln(10)$ 
(which is the value recommended by Bureau Veritas, \cite{bv_2020}). 
\item 
The randomness of $\ln[n(S)]$ is modeled as a Gaussian white noise: i.e., for a given specimen,
the values of $\ln[n(S_1)]$ and $\ln[n(S_2)]$, at two different stress levels $S_1$ and $S_2$, 
are independent random Gaussian variables.
The standard deviation of the white noise is set to $\delta = \delta_{\rm BV}$.
This approach is used for example in \cite{shen_2000, rathod_2011}.
\end{enumerate}
These two scenarios
are illustrated in the right panel of \fig{fig_SNcurve_pattern}.
Computations have shown that option 1 yields the most conservative constraints in the regime 
of small failure probabilities (which is usually the regime relevant for the design of a marine system).
Therefore, it is the modeling option which has been adopted for the case studies presented below.

The fact that option 1 
leads to more conservative constraints compared to option 2
-- in the regime of small failure probabilities --
 can be understood as follows. 
 On the one hand, for a given sequence of stress cycles of various amplitudes, 
 the mean fatigue damage does not depend on the covariance structure of the \sn{} curve.
 On the other hand, the dispersion of the realized fatigue damage (above and below its mean) will be  greater
 when the \sn{} curve randomness is modeled
through assumption 1.
This is because, in option 1, the damages induced at different stress amplitudes are positively correlated,
whereas in option 2,
they are independent.
As a result, option 1 leads to longer tails in the distribution of the realized fatigue damage,
which translates into sizing constraints being more conservative in the regime of small failure probabilities.
The same trend has been found and similarly explained in previous studies 
(see \cite{liu_2007, paolino_2014}), 
where further details are provided.

 \subsubsection{Failure criteria and failure modes}
 \label{subsubsec_fail_crit}

The structure is considered to fail due to fatigue when the cumulated damage, $\Dcum$ (Eq.~\ref{eq_dcum}),
exceeds one.
As the \sn{} curve is randomized through a global shift of the \sn{} curve pattern,
the ultimate strength, $\Su$, 
can be formally identified 
as the stress-cycle amplitude 
at which the realized \sn{} curve intersects $N=1$
(thereby, $\Su$ is a random variable). 
Then, two failure modes can be distinguished:
(i) failure due to a cumulated fatigue damage exceeding one, $\Dcum >1$ 
(with no occurrence of ultimate strength exceedance); 
(ii) failure due to the occurrence of a stress cycle 
with an amplitude exceeding $\Su$.

\subsection{Modeling of sea state occurrence}
\label{subsec_model_st_occurrence}

The long-term sea state history, to which the marine structure is exposed, is modeled as a sequence of short-term stationary sea states.
The randomness of a given stationary sea state is parametrized through its significant wave height, $\hs$ (see \eqp{eq_hs} below), and
its average wave period, $\tz$ (see \eqp{eq_tz} below). 
The assumed sea state population 
is the one recommended by classification societies for the North Atlantic region 
\cite{bv_2019, dnv_2019}.
The related joint density function of sea states in the plane $(\hs,\tz)$,
was derived by applying a conditional modeling approach:
\begin{align}
\pcs(\hs, \tz) = f_{\rm wbl}(\hs; \acs, \bcs, \gcs) \cdot f_{\rm logn}\left[\tz;\mucs(\hs),{\sigcs}^2(\hs)\right] \, ,
\end{align}
where $f_{\rm wbl}$ is the three-parameter Weibull density function of scale parameter $\acs$, shape parameter $\bcs$, and position parameter $\gcs$,
and $f_{\rm logn}$ is the \rom{log-normal} density function of parameters $\mucs$ and ${\sigcs}^2$.
For the North Atlantic region, classification societies recommend 
the following parameter values: 
$\acs = 3.041$, $\bcs = 1.484$, $\gcs=0.661$, $\mucs(\hs)=0.70+1.27 \ {\hs}^{0.131}$ and $\sigcs(\hs) = 0.1334+0.0264\exp(-0.1906 \ \hs)$,
where $\hs$ and $\tz$ are 
in meters and seconds, respectively.
This parametrization is used in this study.

Sea states are assumed to be stationary over a duration of 1h. 
Successive sea states are modeled as independent events.
In practice, an extreme (resp. moderate) sea state, is likely to be followed by another extreme (resp. moderate) sea state.
Accounting for this effect would be important to accurately evaluate the risk of ultimate stress exceedance.
Ignoring the serial dependence of sea states, as done in the present study, 
leads to conservative sizing limits with respect to the risk of extreme event occurrence
(see, e.g., \cite{mackay_2021}).
Conversely, the serial dependence of sea states is unimportant to evaluate the risk of fatigue failure, 
as fatigue damage is built up over a large number of encountered sea states (see \S\ref{subsec_dom_sea_states} for more details). 

\subsection{Modeling of impacts in a given sea state}
\label{subsec_model_ip_occurrence}

The vertical fluid velocity at impact 
is the sole variable used 
to estimate hydrodynamic slamming loads,
and the resulting structural vibratory response (see subsection \ref{subsec_dyn_regimes}).
The impacted body is assumed to be small enough in relation to water wave wavelengths,
so that the body can be reduced to a single material point regarding the risk of slamming event.
Given that ocean waves typically have wavelengths of a few hundred meters,
the latter assumption translates into an upper limit of a few dozen
meters on the horizontal extent of the body.
In this context, 
a slamming event corresponds to the sea surface upcrossing the material point.
For the examples presented below, 
the exposed body is assumed to be fixed in the reference frame of the mean flow 
(i.e., the reference frame where the mean fluid velocity field of water waves is zero).
The way to account for a forward motion, 
and/or a seakeeping motion is exposed in \ref{sec_fwd_sk_motions}. 
In the framework of linear wave theory, the free-surface elevation measured at the station of the material point, 
$\eta(t)$, can be modeled as a Gaussian process,
whose mean is zero (assuming $\eta(t)$ is measured from the mean water level)
and whose power spectrum 
corresponds to the wave spectrum.
The control material point is assumed to be at an altitude $z=a$,
with respect to the mean water level.
In the linear wave model, 
the vertical component of the fluid velocity, $\vz$, at the free surface, 
is equal to the time derivative of the free-surface elevation, $\vz  = {\rm d} \eta / {\rm d} t$.
Using level-crossing theory (see, e.g., \cite{lindgren_2013}), it can be shown that $\vz{|\eta(t) \uparrow a}$,
the vertical velocity given that $\eta(t)$ upcrosses the altitude $\lc$, follows a Rayleigh distribution, 
whose probability density function reads:
\begin{equation}
\label{eq_dist_vz_condi}
f_{\vz{|\eta(t) \uparrow a}} (\vz) = \frac{\vz}{m_2} \exp\left\{ - \frac{1}{2} \frac{\vz^2}{m_2} \right\} \, , \ \vz \ge 0  .
\end{equation}
The scale of the Rayleigh distribution is set by the second moment of the wave spectrum,
\begin{equation}
\label{eq_sw}
m_2 = \int_0^{+\infty} \omega^2 \mathcal{S}(\omega) \ {\rm d} \omega \, ,
\end{equation} 
with $\mathcal{S}$ being the one-sided wave spectrum, 
and $\omega$ being the wave angular frequency.
The significant wave height of the corresponding sea state is defined as:
\begin{equation}
\label{eq_hs}
\hs = 4 \sqrt{m_0}  \, ,
\end{equation}
where $m_0$ is the zeroth moment of the wave spectrum:
\begin{equation}
m_0 = \int_0^{+\infty} \mathcal{S}(\omega) \ {\rm d} \omega   \, .
\end{equation}
The average zero-upcrossing period (i.e., the average wave period), $\tz$, 
is related to $m_0$ and $m_2$ by:
\begin{equation}
\label{eq_tz}
\tz = 2 \pi \sqrt{\frac{m_0}{m_2}} \, .
\end{equation}
By combining Eqs. (\ref{eq_hs}-\ref{eq_tz}), $m_2$ can 
be expressed as a function of the sea state characteristics:
\begin{equation}
\label{eq_sw}
m_2  = \left( \frac{\pi}{2} \frac{\hs}{\tz} \right)^2 \, .
\end{equation} 
In addition to the distribution of impact velocities (\eqp{eq_dist_vz_condi}), 
the frequency of slamming events (i.e., the average frequency of upcrossing events)
is also required
to evaluate the risk of failure 
due to wave slamming.
Using Rice's formula \cite{rice_1944, rice_1945} for a stationary Gaussian process, this frequency can be expressed as:
\begin{equation}
\label{eq_upfreq_gen}
\mu_{\uparrow \lc} = \frac{1}{2 \pi} \sqrt{\frac{m_2}{m_0}} \exp\left\{ - \frac{1}{2} \frac{\lc^2}{m_0}  \right\} \, .
\end{equation}
Finally, by substituting Eqs. (\ref{eq_hs}-\ref{eq_tz}) into \eq{eq_upfreq_gen}, the upcrossing frequency can be expressed as:
\begin{equation}
\label{eq_upfreq}
\mu_{\uparrow \lc} = \frac{1}{\tz} \exp\{ - 8 (\lc/\hs)^2 \} \, .
\end{equation}
Thus, from Eqs.~(\ref{eq_dist_vz_condi}-\ref{eq_sw}-\ref{eq_upfreq}), 
it appears that both the impact frequency and the distribution of impact velocities, for a given sea state, 
are fully determined by the values
of $\hs$ and $\tz$.

\vspace{-0.1cm}
\section{Case studies}
\label{sec_case_studies}
\vspace{-0.05cm}

Building on the framework introduced in the previous section,
the question of whether slamming-induced fatigue can be an important failure mode is now addressed.
To explore this question, a series of
case studies have been investigated 
through numerical simulations.
The considered case studies are described in Section \ref{subsec_sim_configs}.
Section \ref{subsec_num_methods} 
provides a brief overview of
the numerical setup implemented to
conduct the numerical simulations.

\vspace{-0.05cm}
\subsection{Simulated scenarios}
\label{subsec_sim_configs}

The various configurations, considered in this study, 
have been defined by adopting simple assumptions and by varying key parameters, as follows:
\begin{itemize}
\item Long-term exposure duration: the body exposed to wave slamming, is 
assumed to operate continuously for a specific
duration, $\dlt$,
in the same geographical region.
To investigate the effect of exposure duration, 
the following durations have been considered: $\dlt = 1;5;20$ yr.
\item The random encounter of sea states is modeled by following the approach introduced in \S\ref{subsec_model_st_occurrence}.
Sea states are drawn from the North Atlantic population,
assuming that sea states are sequentially independent. 
Each sea state is
assumed to be stationary over a duration $\dst = 1$ hr.
\item The detailed sizing of the structure is 
summarized by the stress-response normalization factor $S_*$, 
introduced in \eq{eq_s1_ss}. 
The damping ratio of vibrations following wave impact (see \eqp{eq_sn_s1}) is fixed at $\zeta = 10^{-2}$.
\item It is assumed that the body is fixed in the reference frame of the mean flow,
at a given altitude, $\lc$, relative to the mean water level
(see Section \ref{subsec_model_ip_occurrence}).
From \eq{eq_upfreq}, the mean number of impacts, for a single sea state exposure, is given by:
\begin{equation}
\label{eq_mean_impact_number}
E[\nipst] = \frac{\dst}{\tz} \exp\left\{ - 8 \left( \frac{\lc}{\hs} \right)^2 \right\} \, ,
\end{equation}
which is used to determine 
the number of impacts generated by each encountered sea state.
In principle, the number of impacts, generated by a given sea state, should be treated as a random variable.
In this study,
$E[\nipst]$ is used as a proxy for $\nipst$.
This simplifying assumption is valid to evaluate the risk of fatigue-induced failure, 
since fatigue damage is built up by a very large number of wave impacts, themselves generated by a large number of encountered sea states.
\item The randomness of the \sn{} curve is modeled according to approach 1, 
as introduced in \S\ref{subsubsec_snRnd}.
\end{itemize}

\subsection{Numerical methods}
\label{subsec_num_methods}

For each case study investigated,
a total number of $\nlt = 10^4$ 
realizations of long-term exposure
are simulated by using a Monte Carlo method.
First, for each long-term realization, a sequence of encountered sea states, 
covering the long-term exposure duration $\dlt$, is drawn in the plane $(\hs, \tz)$.
Then, for each sea state, the number of impacts is computed from \eq{eq_mean_impact_number},
and the set of realized impact velocities is drawn from a Rayleigh distribution whose scale is determined 
by \eq{eq_sw}.
This process yields a long-term collection of impacts, 
from which the long-term collection of stress cycles can be derived by using Eqs. (\ref{eq_sn_s1}-\ref{eq_s1_ss}), 
up to a factor $\sst$ (representative of the structure sizing).
For numerical efficiency, the realized collections of stress-cycle amplitudes are reduced into histograms.
Each long-term histogram of stress-cycle amplitudes is then paired with a realized \sn{} curve 
(i.e., a material sample).
Finally, for a given value of $\sst$, 
the long-term fatigue damage is calculated
by applying Miner's rule (\eqp{eq_dcum}).

 \section{Results}
 \label{sec_results}

\subsection{Probability of failure}
\label{sec_prob_fail}

This section examines the probability of failure 
related to fatigue damage.
\fig{fig_Pfail_Sstar_cmp_ultimate} 
illustrates how
the failure probability, $\pfail$, evolves
as a function of the structural sizing factor, $\sst$.
For a given configuration, 
$\pfail(\sst)$ is an increasing function, from $0$ 
(for $\sst = 0$ there is no structural excitation, and thereby no damage) 
to $1$ (for $\sst \rightarrow \infty$, failure occurs almost surely).
As expected
and shown in \fig{fig_Pfail_Sstar_cmp_ultimate}, 
the probability of failure due to fatigue damage (solid lines in \fig{fig_Pfail_Sstar_cmp_ultimate}) 
rises when the exposure time is increased. 
For a fixed failure probability $ \pfail = \psmall$, 
within the range $\psmall < 0.1$,
increasing the exposure time from $\dlt = 1$ yr to $\dlt = 20$ yr
results in a reduction of
the structural sizing factor, $\sst$, by a factor $\simeq 2.2$.

% %%%%%%%%%%%%%%%%%%%%%%%%
% %%%%%%%%%%%%%%%%%%%%%%%%
% %%%%%        FIGURE    BEGIN         %%%%%
% %%%%%%%%%%%%%%%%%%%%%%%%
% %%%%%%%%%%%%%%%%%%%%%%%%

\def\scaleF{0.43}

\begin{figure}[h!]
\begin{center}
\begin{tabular}{c} 
        \includegraphics[height=\scaleF\textwidth,trim={0 0 0 0.45cm},clip]{\fI 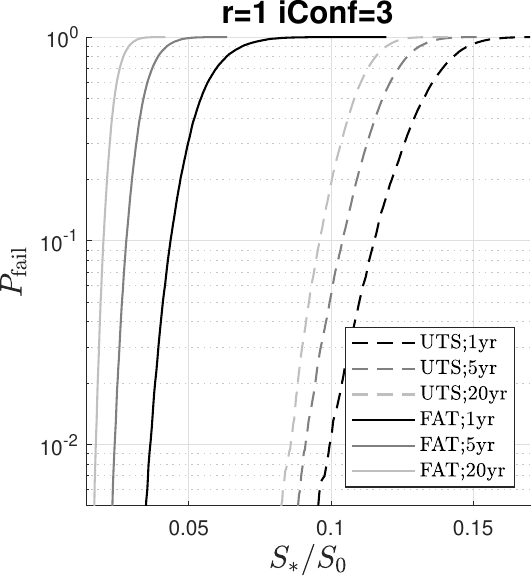} 
\end{tabular}        
\end{center}
\vspace{-0.4cm}
\caption{
Risk of failure, $\pfail$, as a function of the structural sizing factor, $\sst$.
Comparison of $\pfail$ estimates when the risk induced by fatigue is accounted for
(curves labeled as `FAT' in the legend), 
or when only the risk of ultimate strength exceedance is considered 
(curves labeled as `UTS' in the legend).
The elevation of the body is set to $\lc=0$.
Three different exposure durations are considered: $\dlt  = 1 \ {\rm yr}$ (black), 
$\dlt  = 5 \ {\rm yr}$ (dark \rom{gray}),  and $\dlt  = 20 \ {\rm yr}$ (light \rom{gray}).
}
\label{fig_Pfail_Sstar_cmp_ultimate}
\end{figure}

% %%%%%%%%%%%%%%%%%%%%%%%%
% %%%%%%%%%%%%%%%%%%%%%%%%
% %%%%%        FIGURE    END         %%%%%
% %%%%%%%%%%%%%%%%%%%%%%%%
% %%%%%%%%%%%%%%%%%%%%%%%%

\fig{fig_Pfail_Sstar_cmp_ultimate} 
also displays (as dashed lines)
the probability of failure obtained when only the risk of ultimate strength 
exceedance is considered, while ignoring fatigue damage.
Below, 
$\pfailu$ is used to denote the probability of ultimate strength exceedance for a given sizing factor;
and $\sstu$ is used to denote the sizing factor corresponding to a given probability of ultimate strength exceedance.
Note that the estimate of $\pfailu$ is rough,
as the present approach has not been designed to specifically predict extreme events (i.e., extreme waves).
As expected, for a given configuration and a given value of $\sst$, $\pfailu$ is smaller than $\pfail$,
since the risk induced by fatigue is ignored when computing $\pfailu$.
Furthermore,
there is no overlap between the risk related to fatigue damage and the risk related to ultimate strength exceedance.
Here, ``no overlap'' means that -- as $\sst$ is decreased (for a given configuration) -- 
$\pfailu$ becomes vanishingly small before $\pfail$ starts to significantly 
deviate from unity.
Consequently, 
in the region of the parameter space where $\pfail \la 0.1$ the risk of failure is primarily
due to fatigue damage,
while the risk of ultimate strength exceedance brings no significant contribution.
Also, for a given probability of failure, the structural sizing factor 
is noticeably larger (i.e.,  noticeably less conservative), when the risk due to fatigue damage is ignored. 
Among the considered configurations, 
for a small failure probability $\psmall \la 0.03$ (region of the parameter space usually relevant for practical applications), 
the ratio $\sstu(\psmall) / \sst(\psmall)$ is in the range $\simeq 3-5$, depending on the exposure duration.

\subsection{Properties of realizations leading to failure}
\label{sec_rea_to_fail}

This section focuses on \rom{identifying}
the dominant classes of sea states and wave impacts 
that lead to failure.
In the context of reliability analysis, 
a key question arises:
is the damage leading to failure 
due to a large number of 
relatively frequent events, 
or is it driven by rare and extreme events?
Additionally, when considering
the modeling of wave impacts, 
an important related question concerns the degree of nonlinearity of risk-predominant sea states and waves.
These two questions are addressed 
in terms of sea states in \S\ref{subsec_dom_sea_states}, 
and in terms of individual waves in \S\ref{subsec_dom_waves}.
Subsequently, 
\S\ref{subsec_dom_dam_chan} investigates the relative contributions of the different fatigue regimes
to the risk of failure.
These issues are analyzed
for one specific 
exposure duration: $\dlt=10$ yr.
The structural sizing factor $\sst$ is set to a value such that the probability of failure is $\pfail = 2.3\%$,
over the exposure duration.
The elevation of the body is still set to $\lc=0$.
It has been checked that the specific choice of the considered configuration is not crucial 
for the qualitative features 
presented below.

\subsubsection{Dominant class of damaging sea states}
\label{subsec_dom_sea_states}

The left panel of \fig{fig_cumDam_kst}
shows realized curves of cumulated damage as a function 
of the sea state wave steepness, $\kst$,
which is defined by:
\begin{equation}
\label{eq_kst_final}
\kst = \frac{2 \pi}{g} \frac{\hs}{\tz^2} \, ,
\end{equation}
where $g$ is the acceleration due to gravity.
The curves represented are restricted to long-term realizations 
that resulted in failure 
(recall that $\pfail = 2.3\%$ in the present case study);
a total of $\nreaPlot$ realizations resulting in failure are depicted.
To construct each curve, encountered sea states -- until failure occurrence -- were
first collected and sorted by their steepness in descending order.
Therefore, \fig{fig_cumDam_kst} is suitable 
for identifying predominant classes of damaging sea states,
in terms of wave steepness.
The sample of realized curves $\Dcum(\kst)$ shows little dispersion.
Approximately $80\%$ of the cumulated damage is generated by sea states 
with wave steepness values within the range 
$\simeq 0.030-0.072$, i.e., 
a range of relatively ``common'' steepness values.

% %%%%%%%%%%%%%%%%%%%%%%%%
% %%%%%%%%%%%%%%%%%%%%%%%%
% %%%%%        FIGURE    BEGIN         %%%%%
% %%%%%%%%%%%%%%%%%%%%%%%%
% %%%%%%%%%%%%%%%%%%%%%%%%

\def\scaleF{0.43}

\begin{figure}[t!]
\vspace{-0.8cm}
\begin{center}
\begin{tabular}{cc} 
        \includegraphics[width=\scaleF\textwidth]{\fI 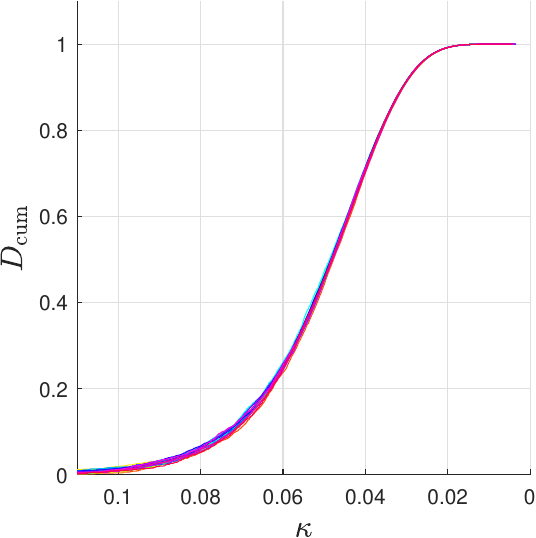} 
        & 
        \includegraphics[width=\scaleF\textwidth]{\fI 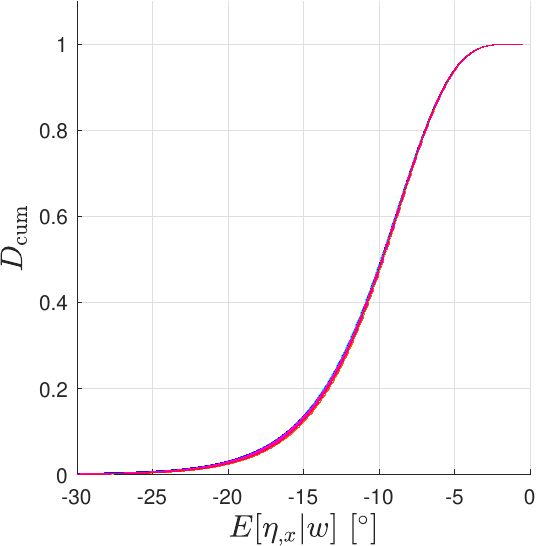} \\ 
\end{tabular}        
\end{center}
\vspace{-0.4cm}
\caption{
\textbf{\textit{Left:}}
Cumulative damage to failure, $\Dcum$, as a function of sea state wave steepness, $\kst$. 
\textbf{\textit{Right:}}
Cumulative damage as a function of the conditional mean of the free-surface slope, $\sx$, 
given the impact vertical velocity, $\vz$, and the sea state properties (see \eqp{eq_condi_slope}).
In both panels the number of represented realizations is $\nrea = \nreaPlot$.
All curves are almost superimposed, 
which shows that the randomness of 
the realized sea-state sequence and wave sequence is
unimportant in the present context.
}
\label{fig_cumDam_kst}
\end{figure}

% %%%%%%%%%%%%%%%%%%%%%%%%
% %%%%%%%%%%%%%%%%%%%%%%%%
% %%%%%        FIGURE    END         %%%%%
% %%%%%%%%%%%%%%%%%%%%%%%%
% %%%%%%%%%%%%%%%%%%%%%%%%

\subsubsection{Dominant class of damaging impacts}
\label{subsec_dom_waves}

Continuing from
the previous subsection, 
this one investigates 
which populations of water waves are predominant in terms of damage,
with a focus on the degree of wave nonlinearity.
To characterize the nonlinearity level of impacting waves, 
the slope of the free surface, at impact, is one quantity of particular interest.
This quantity can be used as a proxy for the steepness of the impacting wave.
\ref{proxy_wave_slopes} demonstrates 
that the conditional mean of the free-surface slope 
-- given the impact velocity, $\vz$, and the sea state properties --
can be used as a reasonable proxy for the realized value of the free-surface slope.
Hereinafter, this conditional mean 
is 
denoted by $E[\sx | \vz]$; its expression is given in \ref{proxy_wave_slopes}, \eq{eq_condi_slope}.
According to
\eq{eq_condi_slope}, $E[\sx | \vz]$ is negative for $\vz > 0$, 
the latter being always satisfied for a free-surface upcrossing event (within the linear wave model).
$E[\sx | \vz]<0$ indicates that the conditional mean slope is downward in the direction of wave propagation.

The right panel of \fig{fig_cumDam_kst}
shows, for a sample of realizations leading to failure,
the cumulative damage, $\Dcum$, as a function of the conditional mean slope, 
$E[\sx | \vz]$. 
To construct each curve, 
the encountered impacts -- until failure -- have been collected and sorted based on their conditional slope,
in descending order of absolute value.
Similarly to the situation observed 
regarding sea state steepness, in \S\ref{subsec_dom_sea_states},
the realized curves of cumulated damage as a function of $E[\sx | \vz]$ exhibit little scatter.
Approximately $80\%$ of the cumulated damage leading to failure is due to waves with $E[\sx | \vz]$ values 
within the range $\simeq -16^{\circ}\rightarrow -6^{\circ}$. 
These slopes can be converted into steepness values of ``equivalent'' regular waves by using the following definition:
\begin{equation}
\label{eq_equi_wave}
k_{\rm reg} A_{\rm reg} \equiv \abs{E[\sx | \vz]} \, ,
\end{equation}
where $k_{\rm reg}$ and  $A_{\rm reg}$ are the wave number and amplitude of the equivalent regular wave, respectively.
The range of slopes $\simeq -16^{\circ} \rightarrow - 6^{\circ}$ corresponds to a range of equivalent-wave steepness
$k_{\rm reg} A_{\rm reg} \simeq (0.089 \rightarrow 0.033) \times \pi$.
This range of steepness values corresponds to 
waves which are moderately nonlinear,
and for which the linear wave model may represent a reasonable approximation.

\subsubsection{Dominant damage channels}
\label{subsec_dom_dam_chan}

This section examines the contribution of the different fatigue damage regimes.
For all realizations leading to failure, the cumulated fatigue damage is comparably distributed between the high cycle and very high cycle regimes,
with $\simeq 39\%$ of fatigue damage occurring in the high cycle regime and $\simeq 61\%$ 
in the very high cycle regime.
 There is virtually no wave impact contributing to fatigue damage in the low cycle regime:
 the probability of such an event occurring is extremely small when $\pfail$ is fixed to a low value ($\pfail = 0.023$ in the present section).

\subsection{Effect of the elevation of the exposed body}
\label{subsec_effect_aLevel}

In the examples presented in 
Sections \ref{sec_prob_fail}-\ref{sec_rea_to_fail},
the control material point -- representing the exposed body --
has been assumed to lie at the mean water level.
Consequently, all waves lead to water-entry events.
In this configuration, it has been found that the risk related to fatigue damage 
leads to sizing constraints that
are significantly more conservative compared to those
obtained from the risk of ultimate strength exceedance.
As the control point is moved away from the mean water level, 
this conclusion will not hold anymore, at some point.
Indeed, as the material point elevation, $\lc$, increases,
water-entry events become less frequent 
and tend to increasingly select high waves and large-$\hs$ sea states 
(see Eq.~\ref{eq_mean_impact_number}, which gives the mean number of slamming events for a given sea state).
If $\lc$ is increased to the point that the mean number of wave impacts become small over the exposure duration,
fatigue is expected to become irrelevant 
to the risk of failure, since the number of cumulated stress cycles will necessarily be limited.
%

% %%%%%%%%%%%%%%%%%%%%%%%%
% %%%%%%%%%%%%%%%%%%%%%%%%
% %%%%%        FIGURE    BEGIN         %%%%%
% %%%%%%%%%%%%%%%%%%%%%%%%
% %%%%%%%%%%%%%%%%%%%%%%%%

\def\scaleF{0.43}

\begin{figure}[b!]
\begin{center}
\begin{tabular}{c} 
        \includegraphics[width=\scaleF\textwidth]{\fI 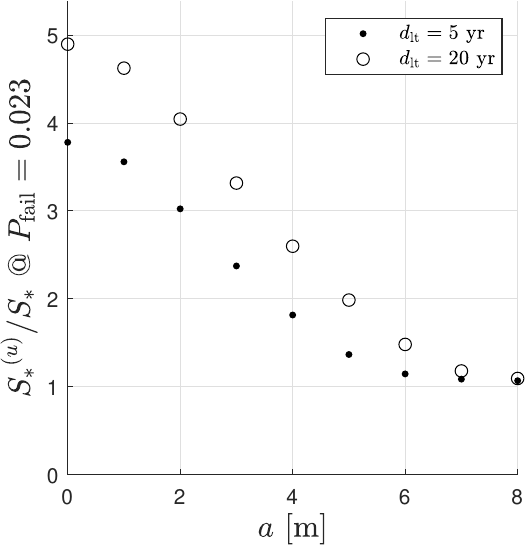} 
\end{tabular}        
\end{center}
\vspace{-0.5cm}
\caption{
Effect of the elevation of the body. 
This figure shows the ratio ${S_*}^{(u)}/S_*$ as a function of the elevation of the body, 
for a failure probability fixed to $\pfail  = 0.023$.
Results are shown for two exposure durations: $\dlt  = 5 $ yr (dots); $\dlt  = 20 $ yr (empty circles).
Further details can be found in the main body of the text.
}
\label{fig_elevation_effect}
\end{figure}

% %%%%%%%%%%%%%%%%%%%%%%%%
% %%%%%%%%%%%%%%%%%%%%%%%%
% %%%%%        FIGURE    END         %%%%%
% %%%%%%%%%%%%%%%%%%%%%%%%
% %%%%%%%%%%%%%%%%%%%%%%%%

%
To apprehend the order of magnitude of $\lc$, 
for which there is a change in the nature of the risk, 
numerical simulations have been 
conducted for different values of $\lc$.
Using the results obtained, \fig{fig_elevation_effect} shows, as a function of $\lc$, the ratio ${\sst}^{(u)}(\pfail)/\sst(\pfail)$, 
for a fixed failure probability, $\pfail = 0.023$.
As a reminder, $\sst(\pfail)$ is the structural sizing factor which includes the risk related to fatigue damage,
and ${\sst}^{(u)}(\pfail)$ is the structural sizing factor obtained  when only the risk of ultimate strength exceedance is accounted for.
\fig{fig_elevation_effect} shows a gradual decrease in ${\sst}^{(u)}/\sst$, as the elevation of the body increases.
For an exposure duration $\dlt = 20$ yr (resp. $\dlt = 5$ yr), 
\fig{fig_elevation_effect} shows that ${\sst}^{(u)}(\pfail)/\sst(\pfail)$ falls below $1.2$ at $\lc \simeq 6.9$ m (resp. $\lc \simeq 5.8$ m).
Note that, as $\lc$ increases, it may become important to account for the effect of wave nonlinearities 
both on the impact frequency (Eq. \ref{eq_upfreq}) and the distribution of the vertical fluid velocity (Eq. \ref{eq_dist_vz_condi}).
This point is discussed in  section \ref{subsec_wave_nonlinearities}, below.

\section{Discussion}
\label{sec_discussion}

Section \ref{subsec_discussion_lin} discusses the results presented above.
Section \ref{subsec_wave_nonlinearities} discusses the anticipated effect of wave nonlinearities, 
and how it may be included in the analysis.
Section \ref{subsec_inference}
briefly explains how the main parameters of the model may be inferred for a specific structural element.
Finally, Section \ref{subsec_discussion_quad} highlights the key challenges 
that would need to be addressed, if the non-impulse regime were to be explored in future research.

\subsection{Relevance of slamming-induced fatigue damage in the impulse regime}
\label{subsec_discussion_lin}

In the impulse regime, the risk of failure due to fatigue 
yields structural-sizing constraints which are noticeably more conservative than those derived solely from 
the risk of ultimate strength exceedance.
In the examples considered in Section \ref{sec_prob_fail},
for a given (small) failure probability, $\pfail$,
the structural sizing factor $\sst$ is $3$ to $5$ times smaller 
(depending on the exposure time)
when fatigue is included in the risk analysis.
The cumulated fatigue damage leading to failure 
mostly occurs in the high cycle and very high cycle regimes.
The probability of a stress cycle reaching the low cycle branch is 
very small and can be considered negligible
 for practical purposes; 
 this conclusion holds for $\lc\la 6 \ {\rm m}$
and $\dlt > 20$ yr.

\subsection{Accounting for wave nonlinearities}
\label{subsec_wave_nonlinearities}

The cumulated fatigue damage is built up by a large number of encountered sea states 
and wave impacts, which are relatively ``common''.
The cumulated damage is mostly due to sea states which have 
characteristic wave steepness values in the range $\kst \simeq 0.030-0.072$.
While
these sea states are not particularly extreme in terms of probability of occurrence, 
the effect of wave nonlinearities on level-crossing statistics may be notable within this range of wave steepness.
Nonlinearities will affect the predicted damage in two ways: 
(i) relative to \eq{eq_upfreq}, 
the frequency of wave impacts will be increased (resp. decreased) for positive (resp. negative) values of the elevation, $\lc$;
(ii) the level-crossing conditioned distribution of wave kinematics (and thus Eq.~\ref{eq_dist_vz_condi}) will be modified.
It is important to note that the extent 
of nonlinearity effects on level-crossing statistics depends on the elevation of the exposed body, $\lc$. 
To the second order, 
the effect of wave nonlinearities on level-crossing statistics remains modest 
as long as $\lc/\hs \ll 1$ \cite{hascoet_2022}.
When $\lc$ is increased and becomes of the same order of $\hs$, 
the effect of wave nonlinearities becomes more notable: 
both the upcrossing frequency, 
and the scale of the upcrossing distribution of the vertical fluid velocity (at the free-surface), increase as $\lc$ increases.\footnote{
Both effects tend to make reliability predictions, based on the linear wave model, unconservative.
} 
A preliminary investigation of the Derisk database \cite{pierella_2021} -- which is populated by numerical simulations based on a fully nonlinear potential code -- 
has shown that these results hold beyond the second order.

Different approaches could be employed to include
wave nonlinearities in the risk analysis.
To the second-order, the effect of wave nonlinearities, 
both on the impact frequency and on the
conditional distribution of wave kinematics, 
may be semi-analytically approached by using Edgeworth series \cite{hascoet_2022}.
Edgeworth series are appropriate for modeling
the bulk of level-crossing distributions,
which is the relevant portion of distributions for fatigue analysis.
Beyond the second order, another 
strategy could involve creating
a library
of upcrossing frequencies and conditional distributions, 
estimated for different sea state conditions.
This library may be populated through experiments (in a lab or in the field),
through simulations, or though a combination of both (using a \rom{multi-fidelity} method).
It should be noted that,
for each sea state condition, 
the experiments (whether physical or numerical)
would not need to be conducted over particularly long time frames, as the
wave impacts that contribute to fatigue damage are not particularly rare events.
Subsequently,
a response model may be fitted to the library, and 
used as a replacement of the first-order wave model, in the long-term analysis.
For instance, the recent database produced within the DeRisk project \cite{pierella_2021} 
may be particularly well suited to produce such a library.

\subsection{On the inference of the main model parameters for a specific structural element}
\label{subsec_inference}

The aim of this study has been to investigate 
whether 
fatigue damage should be considered as a relevant failure mode
when sizing a structural element (e.g., an appendage on a ship, or a tubular element on a platform) 
that locally 
responds to bottom wave slamming. 
Here, ``locally'' applies both to the slamming loads and to the structural response.
A simplified model
has been introduced in Section \ref{subsec_dyn_regimes},
where the body structure is characterized by its
vibratory response time (eigen period of the dominant structural mode), $\tvib$,
and the scale of the stress response to a wave impact, $\sst$ (introduced in Eq.~\ref{eq_s1_ss}).
The curves depicting
$\pfail$ as a function of $\sst$,
 as shown
 in \fig{fig_Pfail_Sstar_cmp_ultimate}, 
 may be used to estimate the probability of failure for a given structural element.
 
 For this purpose, as a first step,
 the characteristic timescales $\tip$ and $\tvib$ need to be estimated,
for the structural element under study. 
The timescale $\tip$ is related to the time evolution of the slamming loads during a wave impact.
It should be evaluated for 
a range of impact velocities
representative of the vertical fluid velocities encountered in water waves (i.e., a few ${\rm m\cdot s^{-1}}$),
from water-entry experiments or numerical simulations.
The vibratory response timescale (i.e., the eigen period
of the dominant structural mode), $\tvib$, may be obtained 
through modal analysis, including the added-mass effect.
Then, by comparing the characteristic times $\tip$ and  $\tvib$, 
the relevant stress-response regime can be identified.
If $\tvib \gg \tip$,\footnote{
See footnote \ref{footnote_harmonic}.
} 
the stress-response is expected to occur in the impulse regime, in which case
the approach developed in 
this study may be applied to assess the risk of failure due to fatigue damage.
In the other cases (i.e., $\tvib \la \tip$), 
a more advanced approach is needed to properly assess the risk of failure (see \S\ref{subsec_discussion_quad}, below).

As a second step, the stress normalization factor, $\sst$, 
needs to be estimated before entering reliability curves as shown in Fig. \ref{fig_Pfail_Sstar_cmp_ultimate}.
For a given detail of the considered structural element, 
$\sst$ may be estimated by measuring the hydroelastic response 
in lab experiments, full-scale (in situ) experiments, or numerical simulations. 
In the case of numerical simulations, high-fidelity modeling may be used, 
since the factor $\sst$ needs to be computed only once 
in the present 
stochastic approach.
Using data from \cite{faltinsen_1999}, 
\ref{fit_sstart_faltinsen} illustrates, through a concrete example, 
how the parameter $\sst$ can be estimated
from numerical simulations conducted at different impact velocities.

Whether $\sst$ should contain stress concentration factors, 
depends on the definition of the stress when entering the \sn{} curve.
The \sn{}  curve pattern used in 
this paper is schematic
and does not assume a specific stress definition.
The design \sn{} curves proposed by classification societies 
depend both on the geometry of the structural detail, 
and on the definition of the stress used to enter the \sn{} curve
(see \cite{bv_2020, dnv_2021a, abs_2017} for details).
It is important to ascertain 
which stress definition should be considered when using \sn{} data.

Experiments or numerical simulations may also be 
valuable for other purposes:
(i) checking that the amplitude of the response depends linearly on the impact velocity (impulse regime),
as formulated in Section \ref{subsec_dyn_regimes},
(ii) confirming the identification of the dominant structural mode, and
(iii) estimating the damping ratio, $\zeta$, of the dominant structural mode.
It may happen that no single mode can be clearly identified as dominant, 
because multiple modes significantly contribute to fatigue damage.
In such a case,
the response may not be modeled through Eqs. (\ref{eq_sn_s1}-\ref{eq_s1_ss}),
and a more advanced approach would be needed.

\subsection{Applicability of the approach to the global response of a vessel}
\label{subsec_discussion_global_response}

Several experimental studies have shown
that slamming events (through the \textit{whipping} mechanism)
may constitute the dominant source of fatigue damage for the hull girder of slamming-prone vessels
(see, e.g., \cite{barhoumi_2014, storhaug_2007c, storhaug_2014, pferdekamper_2024}).
In such a case,
one might wonder whether the formalism introduced in Section \ref{sec_framework} 
is applicable for the sizing of a structural detail whose design stress is directly related to the hull-girder response.
In this perspective, 
the assumptions introduced in Section \ref{sec_framework} would have to be checked.
First, the proposed approach assumes that the impacted body is 
small enough relative 
to water wave wavelengths
that it can be treated as
a single material point regarding the risk of slamming event.
For large vessels, this implies
that the area of the hull experiencing slamming loads
(i) should
be of limited extent, 
(ii) should
be quite repeatable from one slamming event to another.
Another key
assumption is that slamming events can be considered as load impulses,
from the perspective of the vibratory response.
If these conditions are met,
the proposed approach may be applicable.
A critical
experimental test of the applicability of the approach,
would be to check whether the amplitude of the post-slam transient response is linearly correlated to 
the impact velocity (which should consider both wave motion and seakeeping motion),
as formulated in \eq{eq_s1_ss}.
If so, the slope of the correlation between the impact velocity and the amplitude of the stress response of a structural detail 
would yield an estimate for $\sst$.

\subsection{The non-impulse regime}
\label{subsec_discussion_quad}

In this
study, the structural element has been assumed to respond to slamming loads in the impulse regime
($\tvib \gg \tip$).
In the other regimes (i.e., $\tvib \sim \tip$ or $\tvib \ll \tip$), 
a vibratory response may still be excited, even 
when the 
structural response timescale is much shorter than the impact timescale ($\tvib \ll \tip$).
This is because
slamming loads may vary on timescales much shorter 
than the overall impact duration, $\tip$.
For instance, in the case of
blunt bodies (such as a tubular element of circular section), the rise time of slamming loads may be much shorter than the decay time
(see, e.g., \cite{cointe_1987}).
In the regime $\tvib \ll \tip$
(regardless of whether
a vibratory response is excited), 
the magnitude of the stress response is expected to be proportional to the slamming force, and thus
to the square of the impact velocity (for an impact at constant velocity).
This contrasts with
the impulse regime, where the magnitude of stress response 
is proportional to the impact
velocity (see Eq. \ref{eq_s1_ss}).
This change in the velocity dependency exponent increases
the weight of high-velocity impacts in the risk of failure.
A preliminary investigation of this regime 
(results not shown) has been conducted
using the 
same formalism as this study, but with \eq{eq_s1_ss} replaced by:
\begin{equation}
\displaystyle \scyU = \sst \left( \frac{V}{1 \ {\rm m/ s}} \right)^2 \, .
\end{equation}
Compared to the impulse regime, this preliminary investigation 
reveals two key changes 
that significantly complicate the risk assessment:
\begin{enumerate}
\item The low cycle fatigue regime is found to play an important role in the risk of failure.
In this
study, fatigue damage is modeled using Miner's rule. 
A more advanced approach (considering time-history effects) would be necessary 
to accurately estimate fatigue damage in the low cycle regime.
\item Extreme wave impacts are found to play an important role in the fatigue-related risk of failure.
The proper modeling of 
extreme wave impacts (probability of occurrence and wave kinematics) would require a specific 
approach, able to account for
the nonlinearity of the most extreme waves.
\end{enumerate}

\section{Summary and conclusions}
\label{conclusion}

Within the context of the risk of failure induced by bottom wave slamming on marine structures,
the main objective of this
study has been to investigate
whether fatigue damage can be a relevant 
failure mode (i.e., can lead to structural \rom{sizing} being significantly more conservative) compared
to the risk of ultimate strength exceedance.
Without further specifications, the impacted body has been assumed to have a shape and a structural arrangement
such that slamming loads can be considered as impulses from the structural response perspective
(impulse regime).
The sizing of the structure is modeled through a sizing factor, $\sst$,
which sets the magnitude of the stress response, 
assuming 
the latter is dominated by a single vibratory mode.
The \sn{} curve pattern used in this study 
is an extended version of
the fatigue curves recommended by classification societies
for the high cycle and very high cycle regimes.
To combine the risks due to fatigue and due to the ultimate strength exceedance,
the \sn{} curve pattern has been extended into the low cycle regime,
up to the material's ultimate strength.

In the impulse regime, 
fatigue damage is found to be a failure mode important to consider when designing a marine structure.
For a given failure probability, it yields constraints on the sizing factor, $\sst$, 
which can be significantly more conservative 
(by up to a factor $\simeq 5$, in the examples considered in this study)
compared to
the constraints derived solely from the risk of ultimate strength exceedance.
This finding softens as the exposure time, $\dlt$, is decreased and/or the elevation of the body, $\lc$, is increased.

Investigating
the non-impulse regime would require a more advanced approach
that carefully addresses
fatigue damage in the low cycle regime %% is carefully addressed,
and properly models the occurrence of extreme waves. 
This could be the subject of a future study.

% %%%%%%%%%%%%%%%%%%
% %%%%%%%%%%%%%%%%%%
% %%%%%%%%%%%%%%%%%%
% %%%     APPENDICES
% %%%%%%%%%%%%%%%%%%
% %%%%%%%%%%%%%%%%%%
% %%%%%%%%%%%%%%%%%%

% Defining section format
\setcounter{figure}{0}
\appendix

%\newpage
% #########################################
% APPENDIX : Accounting for forward motion and seakeeping motion
% #########################################

\section{Accounting for seakeeping motions and/or forward motion}
\label{sec_fwd_sk_motions}

In the case studies considered above, the material point representing the body 
has been assumed to be fixed in the reference frame of the mean flow.
This appendix explains how the approach can be adapted,
when the structural element under study is attached to a moving platform.
Two types of motion are considered: forward motion and/or seakeeping motions.
Both types of motion are expected to 
affect 
the upcrossing 
frequency (Eq. \ref{eq_upfreq_gen}) and the distribution of water-entry velocities (Eq. \ref{eq_dist_vz_condi}),
but only through changes in
the zeroth and second moments of the wave spectrum. 
With forward motion and seakeeping motion, the moments $m_0$ and $m_2$ appearing in Eqs. (\ref{eq_upfreq_gen}-\ref{eq_dist_vz_condi})
should be replaced with (see \cite{hascoet_2020,hascoet_2021} for details):
\begin{equation}
\label{eq_mTilde_0}
\tilde{m}_0  = \int_{-\pi}^{\pi} {\rm d} \theta \int_{-\infty}^{+\infty} {\rm d} \omega \ 
\abs{1-\mathcal{H}(\omega, \theta; V_s, \psi)}^2 \ G(\omega, \theta) \, ,
\end{equation}
and
\begin{equation}
\label{eq_mTilde_2}
\tilde{m}_2  = \int_{-\pi}^{\pi} {\rm d} \theta \int_{-\infty}^{+\infty} {\rm d} \omega \ 
\left[
\omega - V_s \cos( \psi - \theta ) k(\omega)
\right]^2
\abs{1-\mathcal{H}(\omega, \theta; V_s, \psi)}^2 \ G(\omega, \theta) \, ,
\end{equation}
where the quantities newly introduced are:
\begin{itemize}
\item $\theta$, the direction of wave propagation.
\item $\psi$, the course of forward motion. 
Both $\theta$ and $\psi$ are 
measured relative to a reference direction
attached to the floating platform.
\item $k$, the wavenumber related to $\omega$ through the dispersion relation of water waves,
$
\omega^2 = g \ k \ {\rm tanh}(kh) \, ,
$
with $g$ being the acceleration due to gravity, and $h$ the water depth.
\item $V_s$, the speed of forward motion. The term $\left[ \omega - V_s \cos( \psi - \theta ) k(\omega) \right]$, 
appearing in \eq{eq_mTilde_2}, is the wave encounter frequency, 
induced by the forward motion.
\item $G$, the bidimensional wave spectrum parametrized in terms of wave angular frequency, $\omega$,
and direction of wave propagation, $\theta$.
\item $\mathcal{H}$, the transfer function accounting for the seakeeping vertical motion of the material point. 
The input of the transfer function is a regular wave of frequency $\omega$, and direction of propagation, $\theta$.
The output is the response of the material point (representing the body) in terms of vertical motion.
The term $1-\mathcal{H}$, appearing in Eqs. (\ref{eq_mTilde_0}-\ref{eq_mTilde_2}), 
may be viewed as the transfer function of the free surface elevation measured in the frame of the moving material point.
Using the Response Amplitude Operators of the floating platform
and the position of the structural element on the platform,
the transfer function $\mathcal{H}$ can be readily expressed.
The transfer function $\mathcal{H}$ may also account for the diffraction waves, 
as well as the waves generated by the steady (forward) and oscillatory motions of the platform,
if the related transfer functions are known (see, e.g., \cite{hermundstad_2005}).
\textit{A priori}, the transfer function $\mathcal{H}$ depends on the course and speed of forward motion,
which is why $V_s$ and $\psi$ appear as parameters in Eqs. (\ref{eq_mTilde_0}-\ref{eq_mTilde_2}).
\end{itemize}
The elevation, $\lc$, to be used in \eq{eq_upfreq_gen}, with $m_0$ and $m_2$ replaced by
$\tilde{m}_0$ and $\tilde{m}_2$, 
is the mean elevation of the material point, relative to the mean water level.
For more details about the effect of forward speed on the level-crossing distribution of wave kinematic variables, see \cite{hascoet_2021}.

% #########################################
% APPENDIX : Proxy for the slope of impacting waves
% #########################################

\section{Proxy for impacting-wave slopes}
\label{proxy_wave_slopes}

This appendix shows that the conditional mean of the local free-surface slope 
-- given the impact velocity, $\vz$, and the sea state characteristics, $\hs$ and $\tz$ -- 
can be used as a relevant proxy for the slopes of impacting waves.
For this purpose, a two-dimensional frequency-direction wave spectrum is assumed: 
\begin{equation}
\label{eq_G}
G(\omega,\theta) = \distDir(\theta) \mathcal{S}(\omega) \, ,
\end{equation}
where $G$ is
the two-dimensional wave spectrum, 
$\omega$ the wave angular frequency, and $\theta$ the wave direction of propagation.
In \eq{eq_G}, the frequency and direction dependencies have been assumed to be separable,
with $\mathcal{S}$ being the one-sided wave frequency spectrum, 
and $\distDir$ the wave direction distribution;
the function $\distDir$ satisfies:
\begin{equation}
\int_{-\pi}^{\pi}
 \distDir(\theta) \ \dd \theta  = 1 \, .
\end{equation}
The slope along the 
average wave direction, $\theta = \theta_0$, 
will be denoted as
$\sx$ and it will be used as a proxy for the steepness of the impacting waves.
Within the linear wave model, 
$\sx$ and $\vz$, non-conditioned to level-crossing, are jointly Gaussian,
and are both independent of the free-surface elevation $\eta$ 
(see for instance Section 3.1 in \cite{hascoet_2021}).
Therefore,
the conditional distribution of $\sx$, given $\vz$, is also Gaussian.
The conditional mean and conditional variance are given by:
\begin{equation}
\label{eq_condi_slope}
E[\sx | \vz] = \rho \ssx \frac{\vz}{\sw} \, ,
\end{equation} 
and
\begin{equation}
\label{eq_condi_var}
{\rm Var}[\sx | \vz] = {\ssx}^2 (1-\rho^2) \, ,
\end{equation} 
where $\rho$ is the nonconditional correlation coefficient between $\sx$ and $\vz$,
and $\sw$ (resp. $\ssx$) is the nonconditional standard deviation of $\vz$ (resp. $\sx$).
The nonconditional standard deviation of the vertical fluid velocity is given by $\sigma_{\vz} = \sqrt{m_2}$.
Assuming an infinite water depth, 
the nonconditional standard deviation of $\sx$ is given by (see Section 3.1 in \cite{hascoet_2021}):
\begin{equation}
\label{eq_sig_sx}
{\sigma_{\sx}}^2 = \alpha_{2} \frac{m_4}{g^2} \, ,
\end{equation}
and the correlation coefficient is given by:
\begin{equation}
\label{eq_cov}
\rho \sigma_{\sx} \sigma_{\vz} = E[\sx \vz]  = -\alpha_{1} \frac{m_3}{g} \, .
\end{equation}
In Eqs. (\ref{eq_sig_sx}-\ref{eq_cov}), 
$g$ is the acceleration due to gravity, 
the third- and fourth-order wave moments, $m_3$ and $m_4$, are defined by:
\begin{equation}
\label{eq_spectrum_moment}
m_p = \int_0^{+\infty} \omega^p S(\omega) \ \dd \omega \, ,
\end{equation}
and the coefficients $\alpha_{1}$ and $\alpha_{2}$ depend on the wave direction distribution as follows:
\begin{equation}
\label{eq_alpha_pq}
\alpha_{p} = \int_{-\pi}^{\pi}   \distDir(\theta) \left[ \cos(\theta -\theta_0) \right]^p \ \dd \theta \,.
\end{equation}

Note that $E[\sx | \vz]$ and  ${\rm Var}[\sx | \vz]$
depend on the sea state spectrum 
through $\sigma_{\sx}$, $\sigma_{\vz}$, and $\rho$.
For simplicity,
all sea states are assumed to have the same shape of two-dimensional wave spectrum.
The wave direction distribution is assumed to be of ``cosine-squared'' type:
\begin{equation}
\label{eq_D}
\begin{array}{cc|ll}
\distDir(\theta) & = & (2/\pi) \cos^2 \theta & , \ {\rm for} \ \abs{\theta} < \pi/2 \\
  &  &  0 \,  & , \ {\rm for} \ \abs{\theta} > \pi/2 \,  ,
\end{array}
\end{equation}
with an average wave direction $\theta_0 = 0$.
The wave frequency spectrum is assumed to have a JONSWAP shape \cite{hasselmann_1973}, 
with a peak-enhancement factor $\gamma = 3.3$.
Besides, $1\%$ of the wave energy is truncated at low and high frequency 
(in total $2\%$ of wave energy is discarded).
The high-frequency truncation ensures  that the
variance of the free-surface slope is finite.
The spectrum is normalized following \eq{eq_hs}, \textit{after} the truncation operation.
Eqs. (\ref{eq_sig_sx}) to (\ref{eq_D}) can be used to numerically compute the correlation coefficient:
\begin{equation}
\label{eq_rho_num}
\rho \simeq -0.922 \, .
\end{equation}
The ratio of conditional mean and conditional standard deviation is given by:
\begin{equation}
\label{eq_ratio_mean_var}
\frac{E[\sx | \vz]}{\sqrt{{\rm Var}[\sx | \vz]}} =  \frac{\rho}{\sqrt{1-\rho^2}} \frac{\vz}{\sigma_{\vz}} \, .
\end{equation}
In the impulse stress-response regime,
$\simeq 95\%$ (resp. $\simeq 50\%$) of the cumulated damage is built up by impacts
whose velocities are such that $\vz/\sigma_{\vz}  > 1.4$ (resp. $\vz/\sigma_{\vz} > 2.4$).
Using Eqs. (\ref{eq_rho_num}-\ref{eq_ratio_mean_var}), 
$\vz/\sigma_{\vz}  > 1.4$
corresponds to
$E[\sx | \vz]/\sqrt{{\rm Var}[\sx | \vz]} < -3.3$,
which means that the relative dispersion 
around the mean of $\sx | \vz$ can be considered as moderate.
Therefore,
$E[\sx | \vz]$ can be used as a relevant proxy for the realized slope $\sx | \vz$. 

\section{Estimation of the parameter $\sst$: illustrative example}
\label{fit_sstart_faltinsen}

% %%%%%%%%%%%%%%%%%%%%%%%%
% %%%%%%%%%%%%%%%%%%%%%%%%
% %%%%%        FIGURE    BEGIN         %%%%%
% %%%%%%%%%%%%%%%%%%%%%%%%
% %%%%%%%%%%%%%%%%%%%%%%%%

\def\scaleF{0.55}

\begin{figure}[t!]
\begin{center}
\begin{tabular}{c}        \includegraphics[width=\scaleF\textwidth]{\fI 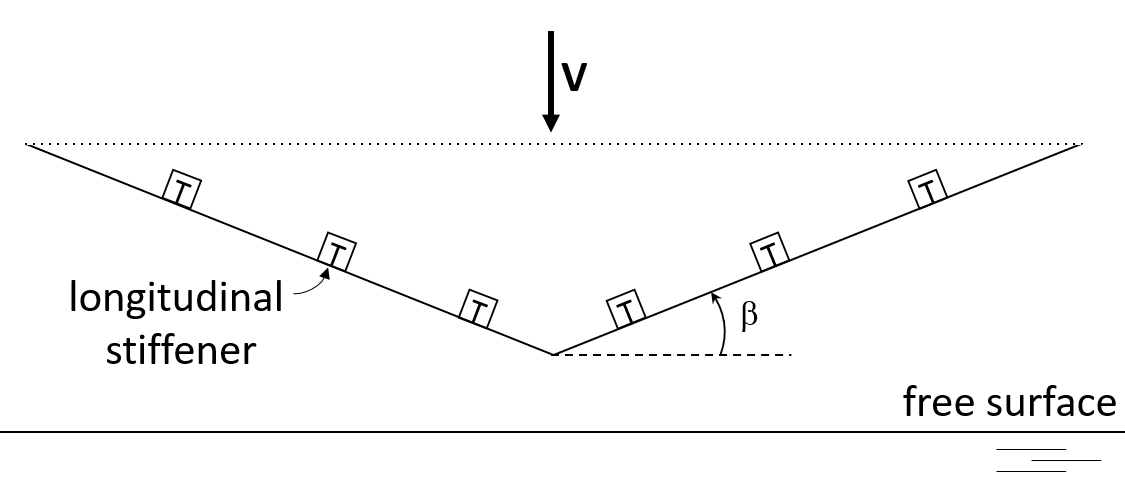} 
\end{tabular}        
\end{center}
\caption{
\rom{
Cross section of the wedge-shaped structure considered in Faltinsen (1999).
The maximum strain reported in \fig{fig_faltinsen} 
is measured in the middle of the second longitudinal stiffener 
(the one pointed to by an arrow on the diagram).
See \cite{faltinsen_1999} for more details.
}
}
\label{fig_wedge}
\end{figure}

% %%%%%%%%%%%%%%%%%%%%%%%%
% %%%%%%%%%%%%%%%%%%%%%%%%
% %%%%%        FIGURE    END         %%%%%
% %%%%%%%%%%%%%%%%%%%%%%%%
% %%%%%%%%%%%%%%%%%%%%%%%%

% %%%%%%%%%%%%%%%%%%%%%%%%
% %%%%%%%%%%%%%%%%%%%%%%%%
% %%%%%        FIGURE    BEGIN         %%%%%
% %%%%%%%%%%%%%%%%%%%%%%%%
% %%%%%%%%%%%%%%%%%%%%%%%%

\def\scaleF{0.45}

\begin{figure}[t!]
\begin{center}
\begin{tabular}{c}        \includegraphics[width=\scaleF\textwidth]{\fI 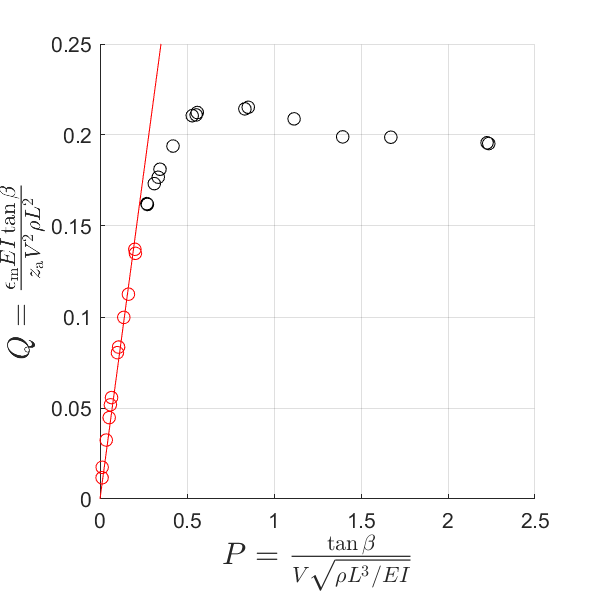} 
\end{tabular}        
\end{center}
\caption{
Nondimensional maximum strain, $Q$, as a function 
of a parameter $P$ that is proportional to $\tip / \tvib$
(see main text).
Numerical data points are shown as circles.
The data points with $P < 0.25$ are highlighted in red: 
they belong to the impulse regime, $\tip \ll \tvib$. 
The red solid line is a least-squares fit of these data points
with a linear function $Q = \Theta P$.
[Data points were extracted from Fig. 17 in \cite{faltinsen_1999}]
}
\label{fig_faltinsen}
\end{figure}

% %%%%%%%%%%%%%%%%%%%%%%%%
% %%%%%%%%%%%%%%%%%%%%%%%%
% %%%%%        FIGURE    END         %%%%%
% %%%%%%%%%%%%%%%%%%%%%%%%
% %%%%%%%%%%%%%%%%%%%%%%%%

This appendix shows how the parameter $\sst$ 
(introduced in Eq. \ref{eq_s1_ss}) can be estimated
in a concrete situation.
The considered case study 
is a wedge-shaped structure 
which was experimentally and numerically investigated 
by Faltinsen (1999) \cite{faltinsen_1999}.
The study examines the importance of hydro-elastic effects 
as a function of wedge deadrise angle and impact velocity.
Using a structural model based 
on hydroelastic orthotropic plate theory,
coupled with a Wager-type water entry model,
the author made numerical predictions of the maximum stress response in the middle of a longitudinal stiffener
\rom{(see \fig{fig_wedge})}.  
The results are \rom{reported} in 
\fig{fig_faltinsen} (see Fig. 17 in \cite{faltinsen_1999} for the original version).
\rom{They} are presented in terms of two nondimensional quantities: 
\begin{equation}
P = \frac{\tan \beta}{V \sqrt{\rho L^3/EI}} \, \text{, on the x-axis, }
\end{equation}
and 
\begin{equation}
Q = \frac{\epm EI \cdot \tan \beta}{\za V^2 \rho L^2} \, \text{, on the y-axis, }
\end{equation}
where $\epm$ is the maximum strain
reached during the impact,
$\beta$ is the deadrise angle of the wedge,
$V$ is the impact velocity (assumed to be constant), 
$\rho$ is the mass density of the fluid, 
$L$ is the span of the wedge (i.e., its length along a generatrix),
$E$ is the Young's modulus of the material,
$I$ is the area moment of inertia of the longitudinal stiffener (and adjacent plate)
about neutral axis, divided by the spacing between longitudinal stiffeners, and $\za$ is the distance from neutral axis to maximum strain location.
It can be shown that the parameter $P$ is proportional to 
the ratio $\tip/\tvib$, $\tip$ being the impact load timescale,
and $\tvib$ being the wet natural period of the longitudinal stiffener
(see \cite{faltinsen_1999} for more details).
In \fig{fig_faltinsen}, for $P \ga 1.4$, 
$Q$ becomes approximately constant (close to $0.2$),
which implies $\epm \propto V^2$: this corresponds to the regime $\tip \gg \tvib$\rom{,}
%that is 
discussed in Section \ref{subsec_discussion_quad}.
For $P \la 0.25$, $Q$ is approximately proportional to $P$, 
which implies $\epm \propto V$:
this corresponds to the impulse regime, 
$\tip \ll \tvib$, described in Section 
\ref{sec_framework}.
The proportionality coefficient 
can be estimated by fitting a linear model $Q=\Theta P$
to the data points in the range $P < 0.25$:
\rom{an ordinary} least-squares \rom{regression} (solid line in \fig{fig_faltinsen}) yields $\Theta \simeq 0.71$.
Since the amplitude of the first stress cycle is equal to the maximum stress in the impulse regime (see \fig{fig_sketch} for an illustration), 
an estimation of the parameter $\sst$, 
as defined in \eq{eq_s1_ss},
is given by:
\begin{equation}
\sst  = \frac{E\epm}{V}= \Theta \za \sqrt{\rho E L / I }
\,.
\end{equation}

\end{document}